\documentclass[5p]{elsarticle}
\usepackage{amsmath,amssymb,mathrsfs,graphicx,epstopdf,placeins,float}
\usepackage{natbib}
\usepackage[pdfstartview=FitH]{hyperref}

 \newcommand{\beq}[1]{\begin{equation}\label{#1}}
 \newcommand{\eeq}{\end{equation}}
 \newcommand{\bea}[1]{\begin{eqnarray}\label{#1}}
 \newcommand{\eea}{\end{eqnarray}}
 \newcommand\figcaption{\def\@captype{figure}\caption}
 \newcommand\tabcaption{\def\@captype{table}\caption}

\begin{document}
\title{n+1 Dimensional Gravity duals to quantum criticalities with spontaneous symmetry breaking}

\date{\today}
\author{Ding-fang Zeng\footnote{dfzeng@bjut.edu.cn,kaizhao@emails.bjut.edu.cn} and  Kai Zhao}
\address{Physics Department, Beijing University of Technology}
\begin{abstract}
 We reexamine the charged AdS domain wall solution to the Einstein-Abelian-Higgs model proposed by Gubser et al as holographic superconductors at quantum critical points and comment on their statement about the uniqueness of gravity solutions. We generalize their explorations from $3+1$ dimensions to arbitrary $n+1$Ds and find that the $n+1\geqslant5$D charged AdS domain walls are unstable against electric perturbations.\\
 \\
 {\it pacs: 11.15.Ex, 04.25.D-,11.25.Tq}
 \end{abstract}
 \maketitle
 \allowdisplaybreaks

\section{Introduction}

The charged AdS domain walls are spaces interpolating two copies of anti-de Sitter space, one of which preserves the abelian gauge symmetry while the other one breaks it. The two sides of the domain walls have different AdS radius. \if 0Similar neutral domain walls, see \cite{dmWtd,dmWreview}for references, are believed to have been formed in the early universe.  also be observed in condensed matter systems, including in superconductors \cite{tdTypeII1,tdTypeII2,tdTypeII3,tdTypeII4}.\fi
In references \cite{holoSC0TwshapePotentialGubserETC}-\cite{holoSC0TstringMtheoryGubserETC}, S. S. Gubser and collaborators proposed that
the quantum critical behavior and the emergent relativistic conformal symmetry in superfluids or superconductivities in strongly coupled gauge theories can be described by charged AdS domain wall solutions in several Einstein-Abelian-Higgs models. These works are mainly concerned with 3+1 dimension gravity theories and provide solutions they think be uniquely determined by the scalar field potential form and double boundary conditions.

The goal of this paper is to generalize these discussions to arbitrary space-time dimensions and study properties of the charged AdS domain-wall. We will first in section 2 point out that the double boundary conditions quotient by \cite{holoSC0TwshapePotentialGubserETC}-\cite{holoSC0TstringMtheoryGubserETC} are questionable and probably exclude the existence of solution families to the relevant dynamic equations. We then in section 3 provide a new ansatz for the AdS domain walls and the corresponding equations of motion. The solution families in both 3 and 4 dimensions are provided also in this section. While in section 5 we study the electric perturbations to the solution and calculate the related electric-transporting coefficients. The last section contains our main conclusions.

\section{About the unique solution of Domain Wall}
This section has two goals. The first is to provide basic ingredients to study the charged AdS domain walls. The second is to discuss the questionable aspects of Ref. \cite{holoSC0TwshapePotentialGubserETC}-\cite{holoSC0TstringMtheoryGubserETC} about the uniqueness of charged AdS domain wall solutions to the relevant equation of motion. \if 0 The ``uniqueness'' of this two references states that, for an Einstein-Abelian-Higgs model system with given potential form such the following \eqref{EMHaction}-\eqref{higgsPotential}, when specifying the ultraviolet asymptotical behavior of the scalar field, the domain wall solution to the system is unique. Our analysis in the following is mainly from technical aspects. For readers who are more willing to catch physics by intuitions, we provide here a reason supporting\fi

First let us provide the basic ingredients of charged AdS-domain wall studyings. Taking the model of Ref. \cite{holoSC0TwshapePotentialGubserETC} as an example
 \bea{}
 S=\frac{1}{16\pi G_N}\int\!\!d^{n+1}\!x\sqrt{-g}\Big(R
 -\frac{1}{4}F_{\mu\nu}F^{\mu\nu}
 \label{EMHaction}\\
 -\overline{D_\mu\psi}D^\mu\psi-V[\psi]\Big)
 ~,~~
 D_\mu\psi=\partial_\mu\psi-iqA_\mu\psi
 \label{minimalCoupling}
 \\
 V[\psi]=-\frac{n(n-1)}{\ell^2}+m^2|\psi|^2+\frac{\lambda}{2}|\psi|^4.
 \label{higgsPotential}
 \eea
 where we have adapted the dimension from $3$ to arbitrary $n$ and used $\lambda$ instead of Ref. \cite{holoSC0TwshapePotentialGubserETC}'s $u$ to denote the scalar field self-coupling constant. Aiming at solutions dual to emergent infrared (IR) conformal symmetries, the authors of Ref. \cite{holoSC0TwshapePotentialGubserETC} set ansatz
 \bea{}
 ds^2=e^{2A}(-hdt^2+d\vec{x}\cdot d\vec{x})+h^{-1}dr^2
 \\
 \psi=|\psi{(}r{)}|,~A_\mu dx^\mu=\Phi(r) dt~~~~~~
 \label{GubserAnsatz}
 \eea
 and require that in the infrared limit $\psi$ sits on the global minimal of $V[\psi]$, i.e. $|\psi|=\sqrt{-m^2/\lambda}$ and $h=1$, $A=r/\ell_\mathrm{ir}$. While the ultraviolet limit of the solution is constrained by the dual field theory to be some specific AdS featured $|\psi|\propto e^{-\Delta_\psi A}$. Reference \cite{holoSC0TwshapePotentialGubserETC} takes $n=3$, so its equations of motion has the form
 \bea{}
 A''=-\frac{1}{2}\psi'^2-\frac{q^2}{2e^{2A}h^2}\Phi^2\psi^2
 \label{GubserEq12}
 \\
  h''+3A'h'=e^{-2A}\Phi'^2+\frac{2q^2}{e^{2A}h}\Phi^2\psi^2
 \label{GubserEq13}
 \\
 \Phi''+A'\Phi'=\frac{2q^2}{h}\psi^2\Phi
 \label{GubserEq14}
 \\
 \psi''+(3A'+\frac{h'}{h})\psi'=\frac{V_{,\psi}}{2h}-\frac{q^2\Phi^2}{e^{2A}h^2}\psi
 \label{GubserEq15}
 \\
 {\psi'}^2h^2+e^{-2A}q^2\Phi^2\psi^2-\frac{h}{2e^{2A}}{\Phi'}^2
 \label{GubserConstraint}\\
 -2A'h'h-6{A'}^2h^2-V[\psi]h=0
 \nonumber
 \eea
Of these equations, the last one is looked as a constraint and the former 4 second order differential equations are solved numerically. The authors state that among the 8 integration constants, (i) one is used up by the constraint \eqref{GubserConstraint}, (ii) six are used up by the infrared-boundary conditions, i.e. as $r\rightarrow-\infty$, (a)$\psi=\psi_\mathrm{ir}$, (b)$\Phi=0$, (c)$A=r/\ell_\mathrm{ir}$, (d)$h=1$, (e)$\Phi{(}r{)}\rightarrow\phi_0e^{(\Delta_\phi-1)r/\ell_\mathrm{ir}}$, (f)$\psi{(}r{)}\rightarrow\psi_\mathrm{ir}+a_\psi e^{(\Delta_{\psi\,\mathrm{ir}}-3)r/\ell_\mathrm{ir}}$, (iii) the last one is determined by ultraviolet boundary condition that $\psi_{r\rightarrow+\infty}\propto e^{-\Delta_\psi A}$. So the solution to Eqs. \eqref{GubserEq12}-\eqref{GubserConstraint} is unique, as long as the ultraviolet scaling dimension of the operator $\mathcal{O}_\psi$ is fixed. There is no solution family parameterized by $a_\psi$.
\begin{figure}[ht]
 \begin{center}
 \includegraphics[scale=1.02]{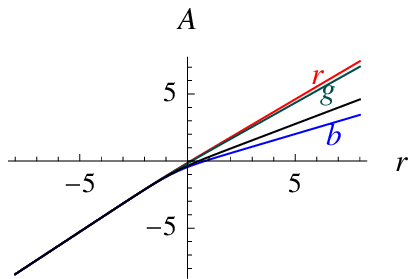}
 \includegraphics[scale=1.02]{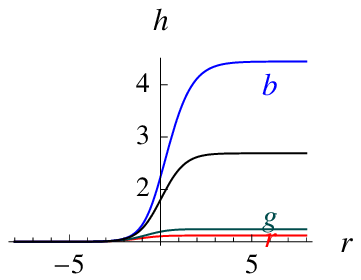}
 \includegraphics[scale=1.02]{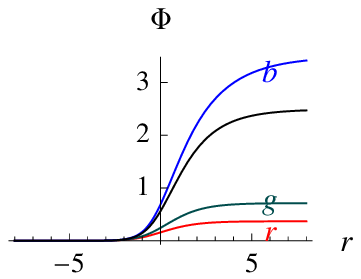}
 \includegraphics[scale=1.02]{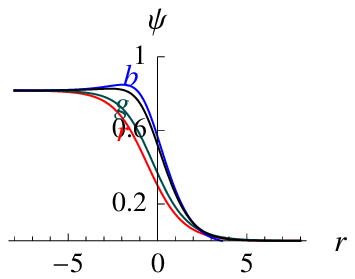}
 \includegraphics[scale=1.02]{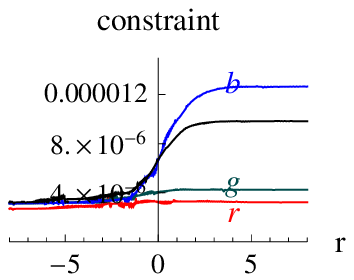}
 \includegraphics[scale=0.5]{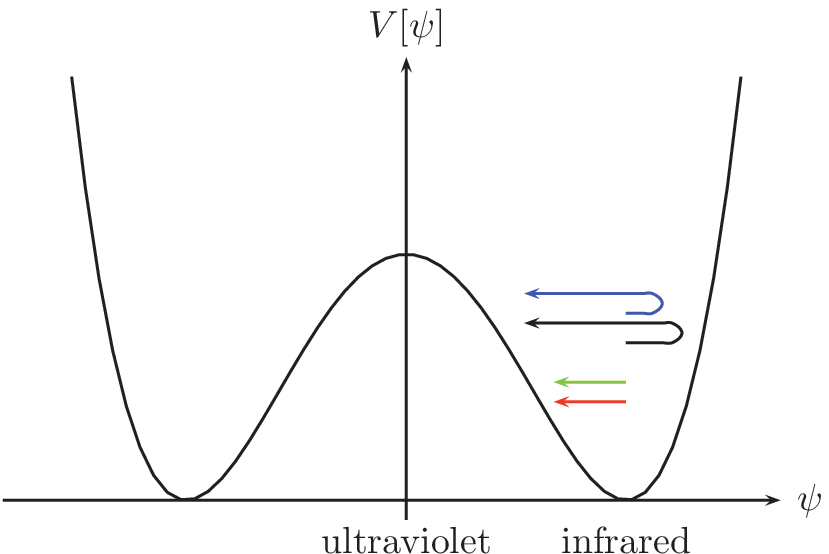}
 \end{center}
 \caption{Domain wall solutions to eqs\eqref{GubserEq12}-\eqref{GubserConstraint}, relevant parameters are chosen the same as \cite{holoSC0TwshapePotentialGubserETC}, i.e. $\ell=1$, $q=2$, $m^2=-2$, $\lambda=3$. In each figures, the lines labeled  r, g, b represent three typical members from a solution family featured by $a_\psi$; the unlabeled black lines represent the `unique' solution of Ref.\cite{holoSC0TwshapePotentialGubserETC}, which is obtained by applying shooting method to a double boundary value problem. In obtaining the three non-unique solutions, we only solve a single boundary value problem where $a_\psi$ are set as $-1$, $-0.5$, $0.5$ freely. On the contrary, in obtaining the `unique' solution by {\it Mathematica}'s NDSolve[$\cdots$,Method$\rightarrow$``Shooting''] command, $a_\psi$'s initial guess is set as $0.3126$ with output $a_\psi=1.62$.
 }\label{figDbcSbc}
 \end{figure}

Here comes our standpoint against unique but supporting family of solutions. We have two reasons for our standpoint. The first is from physical analogue. Just as pointed out by Ref. \cite{fermionNormalModes} and we will emphasize in this paper, the charged AdS domain wall looks on many aspects very like charged extremal black branes \cite{RNBlackBrane}. Obviously, the charge density of extremal black branes can be changed freely as long as its mass density is changed synchronously. For the case of charged AdS domain walls, this is also the case. Changing the charge density of an AdS domain wall just corresponds to changing its chemical potential height and its wall-thickness, see the middle part of figure \ref{figDbcSbc}, where the thickness of the wall can be defined as the characteristic width of the range of $r$ coordinate over which the scalar field varies from $\psi_{ir}$ to $\psi_\mathrm{uv}$. While all these things are implemented by changing the parameter $a_\psi$ of the previous paragraph without changing the asymptotical behavior(equivalence of the conformal dimension of $\mathcal{O}_\psi$) of the scalar field $\psi$. This implies that $a_\psi$ is a freely tunable parameters instead of fixed number determined by the form of scalar field potentials and the conformal dimension of the corresponding field theory operators. This is our first, and probably the most strong reason supporting family instead of unique solutions to the Einstein-Abelian-Higgs model under the AdS-domain wall ansatz.

Our second reason is from technique analysis. In the counting of integration constants consummation of Ref. \cite{holoSC0TwshapePotentialGubserETC}, re-expressed in the second paragraph of this paper, the conditions (a) and (f) are counted repeatedly. Because in the case (b), (c), (d), (e) are all satisfied, the field $\psi$ becomes effectively a scalar field in an AdS space. In this case as long as the value of $\psi$ is set to be $\psi_\mathrm{ir}$, its asymptotic behavior $\psi{(}r{)}\rightarrow\psi_\mathrm{ir}+a_\psi e^{(\Delta_{\psi\,\mathrm{ir}}-3)r/\ell_\mathrm{ir}}$ as $r\rightarrow-\infty$ with $a_\psi$ being an arbitrary constant will be determined exclusively, with no other possibilities. This means that only one of the two conditions (a) and (f) is independent to construct a consistent boundary value problem. Of course, in computer code implementations, to assure that the number of boundary conditions be equal to the number of variables, we still need to write the boundary conditions as
 \begin{subequations}
 \bea{}
 &&\psi(r=r_\mathrm{ir})=\psi_\mathrm{ir}+a_\psi e^{(\Delta_{\psi\,\mathrm{ir}}-3)r_\mathrm{ir}/\ell_\mathrm{ir}}
 \\
 &&\psi'(r=r_\mathrm{ir})=a_\psi(\Delta_{\psi\,\mathrm{ir}}-3)
 e^{(\Delta_{\psi\,\mathrm{ir}}-3)r/\ell_\mathrm{ir}}
 \\
 &&\cdots
 \nonumber
 \eea\label{settingMethod1}
 \end{subequations}
 with $a_\psi$ being set arbitrarily. The fact that conditions (a) and (f) are equivalent implies that we cannot write the boundary conditions as
 \begin{subequations}
 \bea{}
 &&\psi(r=-\infty)=\psi_\mathrm{ir}
 \label{dbc1}
 \\
 &&\frac{\psi'(r=r_\mathrm{ir})}{\psi(r=r_\mathrm{ir})-\psi_\mathrm{ir}}=(\Delta_{\psi\,\mathrm{ir}}-3)
  \label{dbc2}\\
 &&\cdots
 \nonumber
 \eea\label{settingMethod2}
 \end{subequations}
 The setting method \eqref{settingMethod1} accepts that $a_\psi$ is an arbitrary integration constant, while the setting method \eqref{settingMethod2} introduces no any tunable integration constant and try to determine the value of $a_\psi$ through consistences
of the double boundary value problems. We do not know if the author of Ref.\cite{holoSC0TwshapePotentialGubserETC} used setting method \eqref{settingMethod2} but we find that in Mathematica software which uses shooting method to solve double boundary conditions, this setting method can output results indeed! However, when we try to use other algorithms such as relaxations \cite{WHnrbook}, whose principle is translating the double boundary value problem into problem of finding roots to a very large(the size is of the same order as the number of integration steps in solving differential equations) algebraic matrix equations, the boundary setting method \eqref{settingMethod2} always yields singular results. The singularity occurs just due to the fact that two conditions of  \eqref{dbc1}-\eqref{dbc2} are repeatedly used, so the relevant matrix has two columns proportional to each other and not-invertible.

For further supporting the above reasonings, we simultaneously solved the double boundary value problem of \cite{holoSC0TwshapePotentialGubserETC} using shooting method and the single(infrared only, let $a_\psi$ vary freely) boundary-value problem by the usual integration method. To make comparisons, we choose totally the same parameters as \cite{holoSC0TwshapePotentialGubserETC} and reproduce its results faithfully. Our result is illustrated in Fig. \ref{figDbcSbc}. From numerical investigations, we observed the following fact, (i) in reproducing the unique solution of \cite{holoSC0TwshapePotentialGubserETC}, we used the NDSolve[$\cdots$,Method$\rightarrow$``Shooting''] command of {\it Mathematica} software, which requires us to provide an initial guess for parameter $a_\psi$. By general wisdom, its output should not depend on this guess too sensitively. But we find that, this is not the case. For example, as we vary $a_\psi$ from $0.3126$ to $0.312$, warnings appear which tell us that the shooting results may not converge properly. (ii) substituting the output of ``Shooting'' method, $a_\psi=1.62$ into the single-boundary-value problem and re-solving the differential equation, we expect the resulting functions $A{(}r{)}$, $h{(}r{)}$, $\psi{(}r{)}$, $\phi{(}r{)}$ satisfy all differential equations \eqref{GubserEq12}-\eqref{GubserEq15} and the constraint \eqref{GubserConstraint}. But the fact is not, the resulting functions infinitely violate the constraint \eqref{GubserConstraint}, the degree cannot be explained as numerical precision limits. (iii) taking the constraint equation \eqref{GubserConstraint} as the measure of abstract precisions, we see that the ``unique'' solution following from the double value problem is not the one mostly satisfying the constraint equation, instead all members from the solution families with $a_\psi<0$ satisfies the constraint equation more well, see the relevant part of Figure \ref{figDbcSbc}. (iv) in the unique solution of \cite{holoSC0TwshapePotentialGubserETC}, from infrared to ultraviolet region, $\psi$ field first climbs up to the more deeper side of potential well then falls down and then climbs backward to the flatter center extremal, see the last subfigure of figure \ref{figDbcSbc} for intuitions. While in the solutions to the single boundary-value problem, as long as $a_\psi$ is set less than zero, this fact will not occur. Obviously, the climbing-falling-climbing-backward configuration is a more-expensive configuration in field spaces.

Summarizing reasons in this section, we conclude that, given the form of scalar field potentials and the ultraviolet conformal dimension of the corresponding operator $\mathcal{O}_\psi$, there is still a family of charged AdS-domain wall solutions. The members in this family are distinguished from each other by their charged density or wall-thickness. Numerically, it is the parameter $a_\psi$ that determines this features.

\section{New ansatz for the charged AdS domain wall and solutions}
In this section, we introduce a new ansatz for the AdS domain wall and more directly construct the family of solutions. The new ansatz has the advantage of reducing the order of differential equations which follows from minimizing the action of the system,
 \bea{}
 ds^2=e^{2u}(-hdt^2+d\vec{x}\cdot d\vec{x})+\frac{1}{f}du^2
 \label{metricAnsatz}\\
 A_\mu\,dx^\mu=\phi(u)dt~,~~\psi=\psi(u)
 \label{phipsiAnsatz}
 \eea
 As long as $f$ has different values in
 the $u\rightarrow-\infty$ and $u\rightarrow+\infty$ limits, this will
 be domain walls interpolating between two AdS-spaces characterized by $l_{\mathrm{ads}}=1/\sqrt{f_{\mathrm{ir}}}$ and $1/\sqrt{f_{\mathrm{uv}}}$. This ansatz can more explicitly express the asymptotical AdS features of the geometry.
By rescaling $t$ coordinate, we can always set
 $h_{u\rightarrow-\infty}=1$, then the value of $h_{u\rightarrow\infty}$
 will be determined by the equations of motion. As $h$ varies from the infrared region to the ultraviolet region, the speed of light in the two regions will change naturally.
By the ansatz \eqref{metricAnsatz}, the equation of motion reads
 \bea{}
 (n-1)(2n+\frac{h'}{h}+\frac{f'}{f})=-\frac{\phi'^2}{e^{2u}h}-\frac{2V}{f}
 \label{hPf-eom}
 \\
 (n-1)(\frac{h'}{h}-\frac{f'}{f})=2\psi^{\prime 2}+\frac{2q^2\phi^2\psi^2}{e^{2u}hf}
 \label{hOf-eom}
 \\
 \psi^{\prime\prime}+\psi'(n+\frac{1}{2}\frac{h'}{h}+\frac{1}{2}\frac{f'}{f})
 +\frac{q^2\phi^2}{e^{2u}hf}\psi
 -\frac{V_{,\psi}}{f}=0
 \label{psi-eom}\\
 \phi^{\prime\prime}+\phi'(n-2-\frac{1}{2}\frac{h'}{h}+\frac{1}{2}\frac{f'}{f})
 -\frac{2q^2\psi^2}{f}\phi=0
 \label{phi-eom}\\
 \frac{nh'}{h}+\frac{1}{2}\frac{h'f'}{hf}-\frac{1}{2}\frac{h^{\prime 2}}{h^2}+\frac{h''}{h}
 =\frac{\phi^{\prime 2}}{e^{2u}h}
 +\frac{2q^2\phi^2\psi^2}{e^{2u}hf}
 \label{hpp-eom}
 \eea
We checked that in these 5 equations, the last one can be derived out from its four predecessors only by differentiations and combinations. So it is not independent and can be looked as a constraint completely.  Comparing with the equation of motion under the ansatz of Ref.\cite{holoSC0TwshapePotentialGubserETC} and \cite{holoSC0TstringMtheoryGubserETC}, among our four other independent equations, two are first order, while the other two are second order. So essentially have only 6 equivalent first order equations, while the Ref. \cite{holoSC0TwshapePotentialGubserETC} and \cite{holoSC0TstringMtheoryGubserETC} need to solve 8 equivalent first order equations. Obviously, 6 first order differential equations need, and only needs 6 boundary conditions.
 \begin{subequations}
 \bea{}
 &&\hspace{-10mm}h_\mathrm{ir}=1,~
 f_\mathrm{ir}=-\frac{V_\mathrm{ir}}{n(n-1)}=\frac{1}{\ell^2}+\frac{m^4}{n(n-1)2\lambda}
 \label{hfIRbc}\\
 &&\hspace{-10mm}\psi_\mathrm{ir}=\sqrt{-m^2/\lambda}+a_\psi e^{bu},~
 b=-\frac{n}{2}+\sqrt{\frac{n^2}{4}-\frac{4m^2}{f_\mathrm{ir}}}
 \label{psiIRbc}\\
 &&\hspace{-10mm}\phi_\mathrm{ir}=\phi_oe^{ku},~
 k=-\frac{n-2}{2}+\sqrt{\frac{(n-2)^2}{4}+\frac{2q^2\psi_\mathrm{ir}^2}{f_\mathrm{ir}}}
 \label{phiIRbc}
 \\
 &&\hspace{-10mm}
 a_\psi,~\phi_o~\mathrm{can~vary~indepently,~but~one~of~them}
 \label{apsiIRbc}
 \\
 &&\hspace{-10mm}
 \mathrm{can~be~set~to~1~by~shifting~redefinition~of~}u
 \nonumber
 \eea
 \label{IRboundaryConditions}
 \end{subequations}
 The exponent indices $k$ and $b$ involved in these expressions are determined from the infrared limit of equations \eqref{psi-eom} and \eqref{phi-eom}. Note Eq. \eqref{psiIRbc} contains information on two aspects, $\psi$ and $\psi'$ as $u\rightarrow-\infty$, so it should be counted as two boundary conditions.  The same is true for Eq. \eqref{phiIRbc}. The above equations of motion and boundary condition obviously defines a one-parameter family of solutions featured by either $a_\psi$ or $\phi_o$ (our choice is setting $\phi_o=1$ while let $a_\psi$ to feature solutions).
  \begin{figure}[ht]
 \begin{center}
 \includegraphics[scale=0.52]{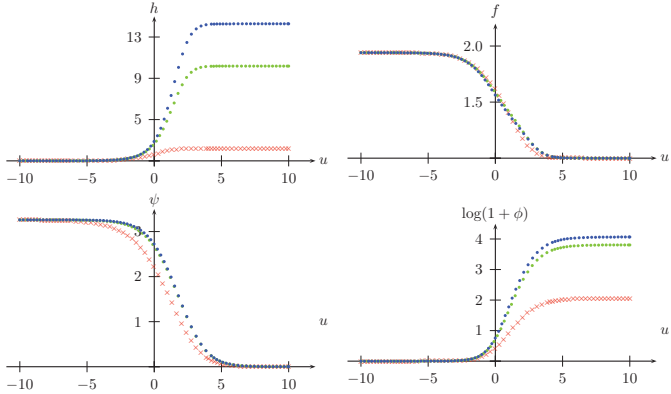}
 \end{center}
 \caption{(color online)Three typical charged domain wall solutions in $n+1=4$ dimensional model. In each part of the figure, from top to bottom, the charge parameters of each curve decrease $a_\psi=-0.01$(blue), $-0.1$(green), $-1$(red)
 respectively. For $n+1=5$ dimensional models, the solutions are similar. The relevant
 model parameters are chosen as
 $\ell=1$, $\lambda=0.2$, $q^2=0.53$, $m^2=-4.25$, numerical
 integration is made from $u=-10$ to $u=10$.
 }\label{figFitSolution1}
 \end{figure}

 Although, the above boundary conditions only specifies the IR behavior of the solution, in the UV limit, as long as $\psi\rightarrow0$, i.e. as long as $\psi$ approaches the meta-stable point of the potential, the fate of other fields, including the asymptotical behavior of $\psi$ itself, are destined. It can be easily proven that
  \begin{subequations}
 \bea{}
 &&\hspace{-5mm}f_\mathrm{uv}=-\frac{V_\mathrm{uv}}{n(n-1)}=\frac{1}{\ell^2},
 ~V_\mathrm{uv}=-\frac{n(n-1)}{\ell^2}
 \\
 &&\hspace{-5mm}\psi_\mathrm{uv}=\psi_o e^{cu},~
 c=-\frac{n}{2}-\sqrt{\frac{n^2}{4}+\frac{2m^2}{f_\mathrm{uv}}}
 \\
 &&\hspace{-5mm}\phi=\phi_\mathrm{uv}+a_\phi e^{pu},~
 p=-(n-2)
 \\
 &&\hspace{-5mm}\phi_\mathrm{uv},~h_\mathrm{uv},~a_\phi,~\psi_o\mathrm{~will~be~determined~by~the}\nonumber
 \\
 &&\hspace{-5mm}\mathrm{IR.b.c~and~eom}
 \nonumber
 \eea
 \label{UVboundaryConditions}
 \end{subequations}
So the key question is, if for various solutions in the family we declared previously, $\psi_\mathrm{uv}$ goes to zero in the ultraviolet limit. Numerics tell us that, this is indeed the case, see Fig. \ref{figFitSolution1}.  Different $a_\psi$ in the infrared limit only leads to different $\phi_\mathrm{uv}$ in the ultraviolet limit and different rate the scalar field $\psi$ evolves from $\psi_\mathrm{ir}$ to $\psi_\mathrm{uv}$. They have no effects on the ultraviolet value of $\psi_\mathrm{uv}$. In the case $\phi_\mathrm{ir}$ is set to zero, the value of $\phi_\mathrm{uv}$ is directly proportional to the charge density of the domain wall. While the rate of $\psi$'s evolution from $\psi_\mathrm{ir}$ to $\psi_\mathrm{uv}$ can obviously be related to the width of the domain wall. From the dual field theory aspect, the charged domain wall describes a finite (but freely variable) density system, the operators dual to $\psi$ have the same conformal dimension. But for different density systems, the evolution of $\mathcal{O}_\psi$'s conformal dimension from $ir$ to $uv$ is different. This is very like the extremal AdS-RN black brane case, whose charge/mass ratio is fixed but the amount of charge and mass each-self are both tunable. The charged AdS domain walls can also carry variable charge densities, but probably fixed charge/mass ratios.

It is worth pointing out that, the difference between heights of different domain walls' electrostatic potentials, see figure \ref{figFitSolution1}, and the corresponding ultraviolet light velocities $h_\mathrm{uv}^{-\frac{1}{2}}$ cannot be tuned away by redefinition of time coordinate. Two things prevent us from doing so. This first is, we have used the redefinition of $t$-coordinate to make $h_\mathrm{ir}=1$. The second is, if the electro-potential height (height in the ultraviolet region relative to that in the infrared region) of a charged domain wall can be changed only through a general coordinate transformation, then we can change the electro-potential height of any such charged domain walls to zero. Obviously, for a charged AdS domain walls, such a fact should not be possible. This discussion suggests us a more closer similarity between the charged AdS domain wall and the extremal AdS-RN black brane. That is, we can change a charged AdS domain wall into a neutral domain wall, just as we change an extremal AdS-RN black brane into a pure AdS-space by reducing their charge and mass density simultaneously.

If we use these charged AdS domain wall systems as models of holographic superconductors, then one possible explanation is that, domain walls with different $a_\psi$ corresponding materials with different charge densities which implement superconductions, i.e. the density of superconductive electrons. The existence of charged domain wall families provides very good examples for the Criticality Paring Conjecture of \cite{holoSC0TstringMtheoryGubserETC}, i.e. (i) in the ultraviolet there is a well defined field theory with the conformal symmetry broken by the finite charge density, (ii) since the finite density deformation in the ultraviolet region, a renormalization flow appears and lead to the broken of continuous symmetries in the infrared region.

 \begin{figure}[ht]
 \begin{center}
 \includegraphics[scale=0.52]{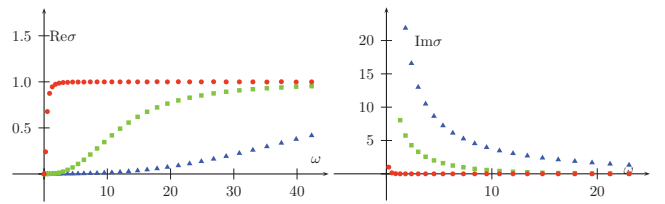}
 \end{center}
 \caption{Conductivity v.s. frequencies relations of the field system dual to
 the $n+1=4$D charged AdS domain walls. In the imaginary part of the figure,
 from top to bottom the charge parameters corresponding to each data set decreases,
 $a_\psi=-0.1$(blue triangle) $-1$(green square) and $-10$(red star)  respectively;
 while in the real part, the order is reversed.
 }\label{figN3conductivity}
 \end{figure}
 \begin{figure}[ht]
 \begin{center}
 \includegraphics[scale=0.52]{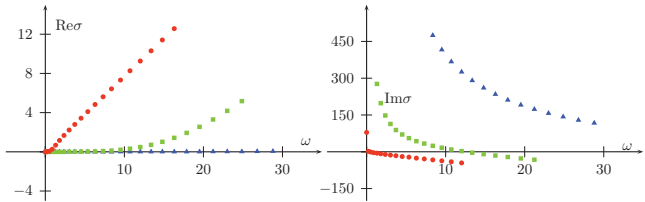}
 \end{center}
 \caption{The same as FIG\ref{figN3conductivity}, but the charged AdS domain walls are $n+1=5$D. Note also that, the imaginary part of $\sigma$ becomes negative as $\omega$ becomes large.
 }\label{figN4conductivity}
 \end{figure}

\section{Electric perturbations and transportations in the dual field theory}
Let us in this section consider the electric perturbations of the charged AdS domain wall solutions. This consideration will give us information on two aspects. One is the stability of the domain wall configuration itself, the other is the transportation properties of the dual field system.

Including responses of the background metric on gauge field perturbations, we can write all the perturbed field as follows,
 \bea{}
 A=\phi(u)dt+a_x(t,x) dx~,~~a_x(t,x)=e^{-i\omega t}a(u)\\
 ds^2=e^{2u}(-hdt^2+d\vec{x}\cdot d\vec{x})+2g_{tx}dtdx+\frac{1}{f}du^2\\
 g_{tx}=e^{-i\omega t}e^{2u}k(u)
 \eea
 From the linearized Maxwell equation and Einstein Equation, we can derive out that, see
 Ref. \cite{HoloSCbackReactHHH}
 \bea{}
 a''+(n-2+\frac{h'}{2h}+\frac{f'}{2f})a'+~~~~
 \nonumber\\
 \Big(\frac{\omega^2e^{-2u}}{hf}
 -\frac{q^2|\psi|^2}{f}-\frac{\phi'^2e^{-2u}}{h}\Big)a=0
 \label{emPerturbEq}
 \eea
 Expanding the solution in the ultraviolet limit in the form
 \beq{}
 a=a_{(0)}+a_{(1)}e^{-(n-2)u}+\cdots
 \eeq
 and using the AdS/CFT dictionary which says that, $a_{(0)}$ is proportional to the perturbing
 source while $a_{(1)}$ to the response i.e. currents in the CFT,
 we directly get the electric conductivity
 \beq{}
 \sigma=\frac{j_x}{E_x}=-\frac{i}{\omega}\frac{a_{(1)}}{a_{(0)}}
 \approx-\frac{i}{\omega}\frac{a'}{a}\frac{e^{(n-2)u}}{-(n-2)}
 \label{conductivity}
 \eeq
Imposing infalling boundary conditions in the infrared limit region and solve equation \eqref{emPerturbEq}, we will get the $\sigma-\omega$ relation directly. Fig.  \ref{figN3conductivity} displays our numerical results for the $n+1=4$  charged AdS domain wall solutions. From
the figure we easily see that, as $|a_\psi|$ decreases, the real part of $\sigma$ decreases correspondingly while the imaginary part increases contrarily. Noting the fact that, smaller $|a_\psi|$ implies higher superconductive-charge density, this is easy to understand from the aspect of two-current model of superconductors. Since in such models, it is the normal electrons (deconfined Cooper pairs) that contribute to the finite part of conductivities $\sigma$. In any given materials, the larger is the density of superconductive electrons, the smaller is the normal ones, and so the smaller is the real part of the conductivities.

By totally the same method of \cite{holoSC0TwshapePotentialGubserETC} and \cite{zeroTSChorowitzETC}, we can verify the scaling law of \cite{holoSC0TwshapePotentialGubserETC} which says that $\sigma$ in small $\omega$ limit is proportional to the simple power of $\omega$,
\beq{}
\mathrm{Re}\,\sigma\propto\omega^{\delta}
~,~~
\delta=2\Delta_{\phi\mathrm{ir}}-1=2k-1
\eeq
When considering the effects of $a_\psi$ on this scaling law, we find that it only modulates the proportional constant in the above relation --- makes it proportional to $h_\mathrm{uv}^{-\frac{\delta}{2}}$. This means
\beq{}
\mathrm{Re}\,\sigma\stackrel{\omega\rightarrow0}{=\!=}
C_{a_\psi}\omega^\delta,~C_{a_\psi}\propto\sqrt{h_\mathrm{uv}(a_\psi)}
\label{powerLawFinding}
\eeq
The last proportionality is also easy to understand, because in the equation governing $a_\psi$'s evolution, $\omega^2/h$ always appears as a whole. By Ref. \cite{holoSC0TwshapePotentialGubserETC}, this scaling law is an indication of quantum criticalities of dual field system. As is well known, strict quantum critical points are transition points occurring at zero temperatures. At such points, the materials manifest two features, conformal symmetry and universal power law correlations. For high $T_c$ superconductor materials, peoples observed that \cite{qcpSuperconductor} there is an optimized doping rate, across which the system manifests obvious power law optical conductivities, which is by definition the two point correlation of electromagnetic currents. Although practical experiments are carried out at finite temperatures, peoples believe that when cooled down to zero temperatures, the high $T_c$ superconducting materials will exhibit strict quantum critical behavior at the optimized doping ratios. So, in high temperature superconductor studies, the value of exploring quantum critical points is, some laws of the high temperature superconductors could probably be results of some expansion around the quantum critical points.

One of the motivations leading references \cite{holoSC0TwshapePotentialGubserETC} and \cite{holoSC0TstringMtheoryGubserETC} to declare the uniqueness of charged domain wall solutions is, they hope to use this model as a description for the quantum criticalities observed in the high $T_c$ superconductors \cite{qcpSuperconductor}-\cite{highTcphasediagram} which occurs only at one optimized doping rate. If the charged domain wall is non-unique, then one must suspect the reasonability of doing this. However, just as the previous section of this work indicates, in a family of charged AdS domain wall solutions, differences between various members are only their charge densities, featured by the parameter $a_\psi$. The observations \eqref{powerLawFinding} tell us that, changing this charge density does not affect the power law feature of the optical conductivity. This implies that, although the quantum phase transition in high $T_c$ superconductors is triggered by optimizing the doping rate of materials, it is not implemented through the changing of superconductive charges' density. In other words, all members in a family of charged AdS domain walls can be used as holographic models of quantum critical superconductors. The only thing worthy of noticing is that, quantum critical superconductions could occur in materials carrying different superconductive charge densities.

Calculating the conductivity of general dimensional ch-\\arged AdS domain wall, we will see that the $n+1\geqslant5$D results are drastically different from those of $n+1=4$D case, see
Fig. \ref{figN4conductivity} and captions there. This difference signifies key properties of higher dimensional charged AdS domain walls. That is, they are unstable as $n+1\geqslant5$. This can also be looked out from the infrared limit analysis of the linearized Maxwell equation \eqref{emPerturbEq}, which says that
 \bea{}
 a''+(n-2)a'+\frac{\omega^2}{h_\mathrm{ir}f_\mathrm{ir}e^{2u}}a=0
 ~,~~\mathrm{or}~~~~
 \nonumber\\
 ~x^2a_{,xx}-(n-3)xa_{,x}+\frac{\omega^2}{h_\mathrm{ir}f_\mathrm{ir}}x^2a=0
 ,~~{e^{-u}\equiv x}
 \eea
where ``$,x$'' denotes derivatives with respect to $x$. The general solution
to this equation reads
 \bea{}
 \hspace{-20pt}\bigg\{\begin{array}{l}
 n=3: a=e^{i{\omega x}/\sqrt{h_\mathrm{ir}f_\mathrm{ir}}}+c\cdot e^{-i{\omega x}/\sqrt{h_\mathrm{ir}f_\mathrm{ir}}}
 \\
 n\geq4:a=x^\nu H^{(1)}_\nu[\frac{\omega x}{\sqrt{h_\mathrm{ir}f_\mathrm{ir}}}]
 +c\cdot x^\nu H^{(2)}_\nu[\frac{\omega x}{\sqrt{h_\mathrm{ir}f_\mathrm{ir}}}]
 \end{array}
 \label{infraredLimitEMPerturbation}
 \eea
 where $\nu=(n-2)/2$. By asymptotic expressions of the Bessel function, we know
 \bea{}
 a\xrightarrow{x\rightarrow\infty}
 x^\frac{n-3}{2}\exp\Big[i\frac{\omega x}{\sqrt{h_\mathrm{ir}f_\mathrm{ir}}}-i\frac{(n-1)\pi}{4}\Big]
 +\nonumber\\
 c\cdot x^\frac{n-3}{2}\exp\Big[-i\frac{\omega x}{\sqrt{h_\mathrm{ir}f_\mathrm{ir}}}+i\frac{(n-1)\pi}{4}\Big]
 \eea
 Obviously, in the $n\geq4$ case, the perturbation
 does not converge. Instead it diverges in the form $x^\frac{n-3}{2}\propto e^{-\frac{n-3}{2}u}$
 as we follow down deep into the infrared region.
 This divergence of the perturbation in the deep infrared region
 obviously implies that, the $n\geq4$ charged domain walls are
 unstable. In the dual field theory, this means that the infrared fixed point (conformal)
 is unstable. From the gravity side, we know this instability neither depends
 on the hight of the domain wall measured by $\psi_\mathrm{uv}-\psi_\mathrm{ir}$
 or $f_\mathrm{uv}-f_\mathrm{ir}$, nor on
 the charge of the domain wall measured by $\phi_\mathrm{uv}-\phi_\mathrm{ir}$
 or $h_\mathrm{uv}-h_\mathrm{ir}$. It is completely determined by the
 dimension of the wall. Although strange, we think this is an interesting result and possibly not being noticed by earlier researchers.

\section{Conclusions}
Two conclusions of this work are worth emphasizing in this section. The first is, given the scalar fields' potential form and its ultraviolet scalings, there is still a domain wall solution family to the relevant equations of motion. Different members of this family carry different charge density but probably fixed charge/mass ratios. Due to differences between the charge densities, different members in this domain wall family have different relative hight of electrostatic potentials in the ultraviolet and infrared region. In the dual field theory, this corresponds to different chemical potential or conserving charge densities. All these things are very similar to the case of extremal RN black holes, whose charge/mass ratio is fixed but the amount of charge or mass each-self is tunable. The second is, the higher dimensional $n+1\geq5$ charged AdS domain wall is perturbatively unstable. Our work uncovers a more closer similarity between the charged AdS-domain wall and the extremal Riessner-Nordstr\"om black branes. That is, the charged AdS-domain wall can be changed into neutral ones just as the extremal AdS-RN black brane can be changed into simple AdS spaces by reducing their charge and mass density simultaneously.

\section*{Acknowledgement}
This work is  supported by Beijing Municipal Natural Science Foundation, Grant. No. Z2006015201001.

 \end{document}